\documentclass{nsart_eng}
\usepackage{cite}
\usepackage[dvips]{graphicx}
\usepackage{amsmath,amssymb}

\year{2022} \volume{0} \nomer{0} \firstpage{1}

\begin{document}

\title[]{Double ring polariton condensates with polariton vortices}

\author[]{$^{1,2,3,4}$E.\,S.\,Sedov,
$^{3,5}$V.\,A.\,Lukoshkin, $^{3,5}$V.\,K.\,Kalevich, $^{6,4}$I.\,Yu.\,Chestnov, $^{7}$Z.\,Hatzopoulos, $^{1,2,6,8}$P.\,G.\,Savvidis, $^{1,2,3,9}$A.\,V.\,Kavokin}

\address{$^1$Key Laboratory for Quantum Materials of Zhejiang Province, School of Science, Westlake University, 18 Shilongshan Rd, Hangzhou 310024, Zhejiang, China\\
$^2$Institute of Natural Sciences, Westlake Institute for Advanced Study, 18 Shilongshan Road, Hangzhou, Zhejiang Province 310024, China\\
$^3$Spin Optics Laboratory, St. Petersburg State University, Ulyanovskaya 1, St. Petersburg 198504, Russia\\
$^4$Vladimir State University, 87 Gorky str. 600000, Vladimir, Russia\\
$^5$Ioffe Institute, Russian Academy of Sciences, 26 Politechnicheskaya, 194021 St-Petersburg, Russia\\
$^6$Department of Physics and Technology, ITMO University, St. Petersburg, 197101, Russia\\
$^7$FORTH-IESL, P.O. Box 1527, 71110 Heraklion, Crete, Greece \\
$^8$ Department of Materials Science and Technology, University of Crete, P.O. Box 2208, 71003 Heraklion, Crete, Greece\\
$^9$Moscow Institute of Physics and Technology, Institutskiy per., 9, Dolgoprudnyi, Moscow Region, 141701, Russia
}

\email{evgeny\_sedov@mail.ru}


\begin{abstract}
We study formation of persistent currents of exciton polaritons in annular polariton condensates in a cylindrical micropillar cavity under the spatially localised nonresonant optical pumping.
Since polariton condensates are strongly nonequilibrium systems, the trapping potential for polaritons, formed by the pillar edge and the reservoir of optically induced incoherent excitons, is complex in general case.
Its imaginary part includes the spatially distributed gain from the pump and losses of polaritons in the condensate.
We show that engineering the gain-loss balance in the micropillar plane gives access to the excited states of the polariton condensate.
We demonstrate both theoretically and experimentally formation of vortices in double concentric ring polariton condensates in the complex annular trap potential.
\end{abstract}


\maketitle

\section{Introduction}

Exciton polaritons are hybrid bosonic quasiparticles emerging under the strong coupling of photons and excitons.
The coupling can be provided in specially designed stratified structures, semiconductor optical microcavities with embedded quantum wells~\cite{kavokinBook2017}.
Microcavities provide localization of light along one spatial direction, reducing the problem to two dimensions.
Quantum wells are holders of excitons close to the resonance with the microcavity photons.

Cavity polaritons are able to form Bose-Einstein condensates.
Such states are known for their superfluid properties~\cite{NatPhys6527,RevModPhys85299}.
Superfluidity implies the existence of persistent polariton currents in the system~\cite{PhysRevLett119067406}.
The configuration of the current field considerably depends on the geometry of the localizing potential for polaritons.
Azimuthal persistent currents have been predicted and observed in polariton condensated trapped in annular potentials~\cite{PhysRevB97195149,ACSPhoton71163,PhysRevResearch3013072,SciRep1122382, PhysRevResearch3013099}.
Such currents can exhibit both vortex and non-vortex nature.
To distinguish them, one needs to pay attention to their phase.
For vortices, phase rotates around the trap by $2 \pi m$, where the integer number $m$ is known as the winding number, which plays the role of the topological charge of the vortex~\cite{NaturePhysics4706}.
The second distinctive feature of vortices, which is the vanishing density of the superfluid at the core, is not relevant for the annular geometry of the potential.

In the above mentioned papers~\cite{PhysRevB97195149,ACSPhoton71163, PhysRevResearch3013072,SciRep1122382}, the polariton condensates supporting circular polariton currents have been characterized by a single ring shape occupying the ground radial state of the annular trap.
As one has shown~\cite{PhysRevB91045305,PNAS1118770}, due to the nonequilibrium nature of polariton condensates, which can exist only under the external (optical) pumping, they are able to occupy the excited states, including those in the form of concentric rings.
Herewith circular polariton currents are also allowed in such condensates~\cite{LukoshkinPaper2022}.

In this manuscript, we consider the system of a cylindrical micropillar cavity excited by a non-resonant laser pump focused close to its center.
The pump creates a reservoir of incoherent excitons, which feeds the exciton polariton condensate through stimulated scattering processes.
We study the conditions needed for exciting condensates in the form of concentric rings in the annular trapping potential by analysing the balance of gain of the condensate from the optical pump and losses of polaritons.
We predict formation of double concentric ring polariton vortices and confirm our predictions by experimental observations.
We note that double ring polariton condensates may be used as building blocks in polariton quantum networks~\cite{NatPhysRev4435}.

\section{Exciton polariton condensates in an annular trap}

\subsection{Gross-Pitaevskii equation for describing polariton condensates}

The traditional approach to characterizing the exciton polariton condensate is based on using the generalized Gross-Pitaevskii equation for the polariton wave function (WF)~$\Psi (t,\mathbf{r})$~\cite{NJPhys14075020,PhysRevLett109216404,CherbuninPaper2022,LukoshkinPaper2022}:
\begin{equation}
\label{EqGPE}
i \hbar \partial _t \Psi =\left\{ [i \eta n_{\text{R}} - 1] \frac{\hbar ^2}{2 M} \nabla ^ 2\right.
 + V (r) 
\left.  +\alpha |\Psi|^2 + \alpha _{\text{R}} n_{\text{R}}  
+ \frac{i \hbar}{2} \left[ R n_{\text{R}} - \gamma \right] \right\} \Psi ,
\end{equation}
where $M$ is the effective polariton mass in the microcavity plane,
$V(r) = V_0 \delta (r - d/2)$ is the stationary potential of the micropillar, taken in the form of cylinder of a diameter $d$ and height $V_0$.
$\alpha$ and $\alpha _{\text{R}}$ are the polariton-polariton and polariton-exciton interaction constants, respectively.
The rightmost imaginary term in Eq.~\eqref{EqGPE} characterizes non-conservative processes of polariton losses and gain from the incoherent reservoir of optically induced excitons.
$n_{\text{R}} (t,\mathbf{r})$ is the density of the reservoir,
$R$ is the stimulated scattering rate from the reservoir to the condensate,
$\gamma$ and $\gamma_{\text{R}}$ are the decay rates of polaritons and excitons, respectively.
In Eq.~\eqref{EqGPE} we also take into account the energy relaxation of polaritons during their propagation~\cite{NJPhys14075020,PhysRevLett109216404}.
$\eta$ is the energy relaxation constant.
This latter term distinguishes our model from one proposed in~\cite{PhysRevResearch3013072}.

Due to larger effective masses and higher losses of excitons, the reservoir dynamics is significantly faster than one of polaritons. 
This allows us to treat the change of the reservoir as the adiabatic adjustment to the evolution of the condensate:
\begin{equation}
\label{EqnRExpanded}
n_{\text{R}} \simeq \frac{P(\mathbf{r})}{\gamma _{\text{R}} + R|\Psi|^2} \approx \frac{P(\mathbf{r})}{\gamma _{\text{R}}} - \frac{P(\mathbf{r}) R|\Psi |^2}{ \gamma _{\text{R}}^2}.
\end{equation}

The reservoir is excited by the optical pump of the intensity $P(\mathbf{r}) = P_{\text{s}} (r) + \delta P(\mathbf{r})$, where  $P _{\text{s}}(r) \propto \exp \left[ - r^2 /2 w^2 \right]$ is the  Gaussian function of width~$w$ centered in the micropillar, and $\delta P(\mathbf{r})$ describes perturbation due to the displacement of the pump spot or deformation of its shape.

\subsection{Gain-loss balance for emergence of the polariton condensate}

Let us start our consideration with the rotationally symmetric ($\delta P(\mathbf{r}) \rightarrow 0$) linear (${\alpha |\Psi|^2 \rightarrow 0}$) system.
The polariton WF then can be factorized as follows: $\Psi _{m,n} = \Upsilon _{n} (r) \exp \left[ - i (E _{m,n} t/ \hbar - m \theta)\right]$, where $\Upsilon _{n} (r)$ is the radial component of WF with the quantum number~$n$. $m$ is the azimuthal quantum number, that is also the winding number characterizing circulation of polaritons around the trap.
$E _{n,m}$ is the complex eigenvalue of the problem, obeying the equation
\begin{multline}
\label{EqSymmRadEigprobl}
\left( E _{n,m} + i\frac{\hbar }{2} \gamma \right)\Upsilon _n 
= \left\{ \frac{\hbar ^2 }{2 M}\left( i\frac{\eta P_{\text{s}}(r)}{\gamma _{\text{R}}} - 1\right) \left( \frac{\partial ^2 }{\partial r^2} + \frac{1}{r} \frac{\partial}{\partial r} - \frac{m^2}{r^2}  \right)  
  \right. \\ 
\left. + V (r) + \frac{ P_{\text{s}}(r)}{\gamma_{\text{R}}} \left[ \alpha _{\text{R}} + i\frac{ \hbar R}{2} \right] \right\} \Upsilon _n.
\end{multline}
The emergence of the condensate is determined by the overall balance of gain and loss in the system.
The gain-loss balance for the state $\Psi _{n,m}$ at a given pump power is determined by the imaginary part of the eigenvalue~$E _{n,m}$.
The pump power, for which the condition ${\text{Im} [E_{n,m}] = 0}$ is satisfied, is the threshold power for the state $\Psi _{n,m}$~\cite{PhysRevB103115309}.
Those states, for which gain exceeds losses (${\text{Im} [E_{n,m}] > 0}$), are supported by the system, and they can be occupied by the polariton condensate.

\begin{figure}[htbp]
\centerline{\includegraphics[width=4in]{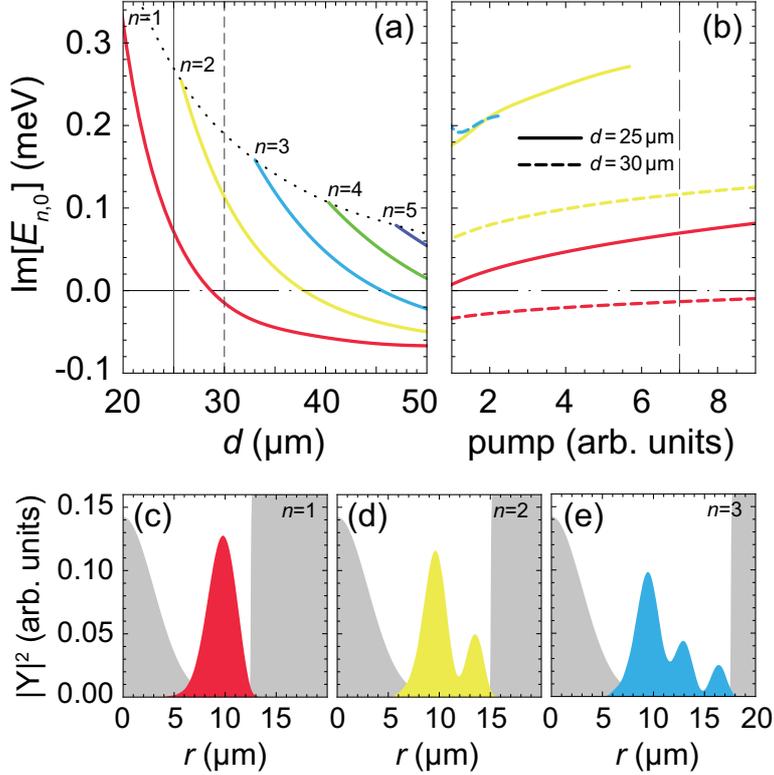}}
\caption{The dependence of the imaginary part of the eigenvalues $E_{n,0}$ for several lower states on the diameter of the pillar, $d$, at the fixed pump power (a) and on the pump power at the fixed diameter (b).
(c--e) Examples of the density distribution of the three lowest eigenstates with $n=1,2,3$.
The vertical solid and dashed lines in (a) as well as the solid and dashed curves in (b) correspond to $d=25 \, \mu \text{m}$ and $30 \, \mu \text{m}$, respectively.
The vertical dashed line in (b) indicates the pump power used for (a).
The gray shaded regions in (c--e) indicate the effective trapping potential for polaritons.
The colors in all panels indicate the number~$n$ of the energy level.} \label{FIG_Radial}
\end{figure}

The gain and loss terms can be included in the effective potential of polaritons, which then acquires a complex form. 
Tuning the parameters of the complex potential, one can manipulate by the emerging polariton condensate modes. 
We have numerically studied the effect of the width of the annular potential and the pump power on the condition of emergence of excited polariton condensate states, see illustrations in~Fig.~\ref{FIG_Radial}.
For numerical simulations, we take the following values of the parameters.
The effective mass of polaritons is $M = 3 \cdot 10^{-5} m_{\text{e}}$, where $m_{\text{e}}$ is the free electron mass.
The height of the stationary potential is $V_0 = 1 \, \text{meV}$.
The polariton and exciton decay rates are $\gamma = 0.1$~ps$^{-1}$ and $\gamma _{\text{R}} = 0.33$~ps$^{-1}$, respectively.
The stimulated scattering rate is $\hbar R=0.1\, \text{meV} \, \mu \text{m}^2$.
The nonlinearity coefficients are $\alpha = \alpha_{\text{R}}/2 = 3 \, \mu \text{eV} \, \mu \text{m}^2$.
The fitting parameter of the energy relaxation is $\eta = 0.01 \, \mu \text{m}^2$.
The pump width is $w = 3 \, \mu\text{m}$.
The pump power simultaneously contributes both the real part of the trapping potential, affecting height of the potential hill from the exciton reservoir, and the imaginary part, affecting gain of polaritons in the condensate.
We would also like to note that although in this manuscript we consider the effective trapping potential which includes the stationary potential of the micropillar, our discussion can be expanded to the case of a fully optically induced potential~\cite{PhysRevB91195308,PhysRevB97235303}.

The dependence of the imaginary parts of the eigenvalues $E_{n,0}$ on the width $d$ of the potential ring at the fixed pump power is shown in Fig.~\ref{FIG_Radial}(a). 
One can see that at a small radius, the system supports only the ground polariton condensate state with ${n=1}$ (see Fig.~\ref{FIG_Radial}(c)), characterized by a single ring shape.
This is realized, e.g., in 25~$\mu \text{m}$ wide micropillars (solid vertical line on the panel) studied in Refs.~\cite{ACSPhoton71163, PhysRevResearch3013072,SciRep1122382}.
At some~$d$ the eigenspectrum of the trap is supplemented with the first excited states with ${n=2}$ (see Fig.~\ref{FIG_Radial}(d)), which, however, remains the only eigenstate of the system with further increase in~$d$ (when $\text{Im}[E_{1,0}]$ becomes negative).
These states characterized by a double concentric ring shape in a 30~$\mu \text{m}$ wide micropillar (dashed vertical line in the panel~(a)) are in the focus of our present consideration.
The further increase in~$d$ brings states with larger~$n$ into play, while the states with smaller $n$ are eliminated.
Similar turnover of the eigenstates takes place with the change of the pump power.
In Fig.~\ref{FIG_Radial}(b) the dependencies of $E_{n,0}$ on the pump power for $d=25 \,\mu \text{m}$ and $30 \,\mu \text{m}$ wide micropillars are shown.

\subsection{Polariton voirtices in concentric ring condensates}
We now return to the general case of the system and bring the pump perturbation, $\delta P(\mathbf{r})$, and polariton-polariton interactions, $\alpha |\Psi|^2$, back into consideration.
We use the approach developed in~\cite{PhysRevResearch3013072} and project Eq.~\eqref{EqGPE} onto the solution $\Upsilon _2 (r) $ of Eq.~\eqref{EqSymmRadEigprobl}.
We substitute the decomposition $\Psi (t, \mathbf{r}) = \Upsilon _2 (r) \Phi (t,\theta)$ in Eq.~\eqref{EqGPE} and after averaging over $r$ we obtain the following equation for the azimuthal WF component~$\Phi (t,\theta)$:
\begin{equation}
\label{EqAzimEq}
i \partial _t \Phi = \left\{ 
\left[ A_1 (\theta) - A_2 (\theta) |\Phi|^2 \right]  \frac{\partial ^2}{\partial \theta ^2} 
+ U_1(\theta) 
+ U_2(\theta)  |\Phi|^2 \right\} \Phi,
\end{equation}
where
\begin{subequations}
\label{EqAveragedParams}
\begin{eqnarray}
&&A_1 (\theta) =  - \frac{\hbar }{2 M} \left\langle \frac{1}{r^2} \left(1 - i\frac{\eta P(\mathbf{r})}{\gamma _{\text{R}}} \right) \right\rangle, \\
&&A_2 (\theta) = -i \frac{\hbar \eta R}{2M \gamma _{\text{R}}^2} \left\langle \frac{1}{r^2} P(\mathbf{r})  |\Upsilon|^2 \right\rangle, \\
&&U_1 (\theta) = 
\frac{i \hbar \eta}{2M \gamma _{\text{R}}} \left\langle \delta P(\mathbf{r}) \left( \frac{\partial ^2}{\partial r^2} + \frac{1}{r} \frac{\partial}{\partial r} \right) \right\rangle + \frac{1}{\gamma _{\text{R}}} \left( \frac{\alpha _{\text{R}}}{\hbar} + i \frac{R}{2} \right) \left\langle \delta P(\mathbf{r})  \right\rangle, \\
&&U_2 (\theta) = - \frac{i \hbar \eta R}{2 M \gamma _{\text{R}}^2}\left\langle P(\mathbf{r}) |\Upsilon|^2 \left( \frac{\partial ^2}{\partial r^2} + \frac{1}{r} \frac{\partial}{\partial r} \right)  \right\rangle
\nonumber \\
&& \,
\phantom{U_2 (\theta) =} - \frac{R}{\gamma _{\text{R}}^2}\left( 
\frac{\alpha _{\text{R}}}{\hbar } + i\frac{R}{2}
\right) \left\langle P(\mathbf{r}) |\Upsilon|^2  \right\rangle + \frac{1}{\hbar} \alpha \left\langle |\Upsilon|^2 \right\rangle. 
\end{eqnarray}
\end{subequations}
In~\eqref{EqAveragedParams}, $\langle ... \rangle = \int _0 ^{\infty} ...  |\Upsilon|^2 r dr$.
We omitted the subscript ``$_2$'' at $\Upsilon$ for brevity.
The functions of angle $U_1(\theta)$ and $U_2(\theta)$ contribute to the symmetry breaking due to the pump perturbation and interaction-induced nonlinearity, respectively. 

As one has shown in~\cite{PhysRevResearch3013072}, breaking the rotational symmetry of a single ring polariton condensate, e.~g., due to weak ellipticity of the pump spot and its small displacement from the center of the micropillar, can lead to the emergence of azimuthal polariton currents.  
We apply similar approach to the concentric ring geometry of the condensate.
We take the pump in the form $P(\mathbf{r}) \propto \exp \left[\left. -(\mathbf{r} - \mathbf{r}_{\text{p}})^2 \right/ (1+s \cos^2 \theta )w^2  \right]$, where $s$ is responsible for  deformation of the pump spot, $\mathbf{r}_{\text{p}}$ describes the displacement of the pump spot.
Keeping $s$ fixed and solving Eqs.~\eqref{EqSymmRadEigprobl}--\eqref{EqAzimEq} with different~$\mathbf{r}_{\text{p}}$, we simulate concentric ring polariton condensates with and without azimuthal polariton currents.

\begin{figure}[htbp]
\centerline{\includegraphics[width=4in]{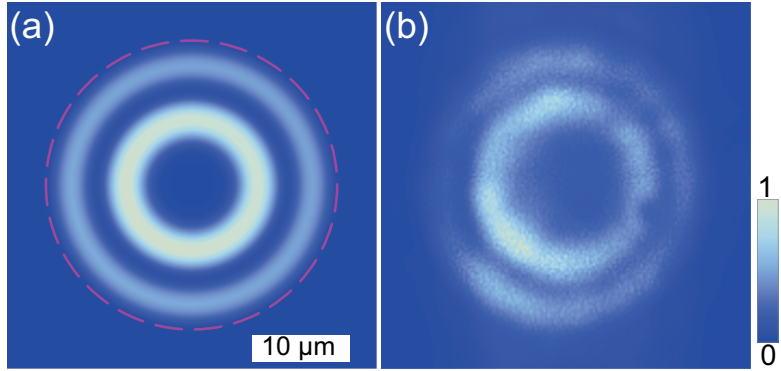}}
\caption{Simulated (a) and experimentally observed (b) density distribution of the polariton condensate in the shape of double concentric rings.
The magenta dashed circle in (a) indicates the edge of the micropillar.} \label{FIG_Density}
\end{figure}

\begin{figure}[htbp]
\centerline{\includegraphics[width=4in]{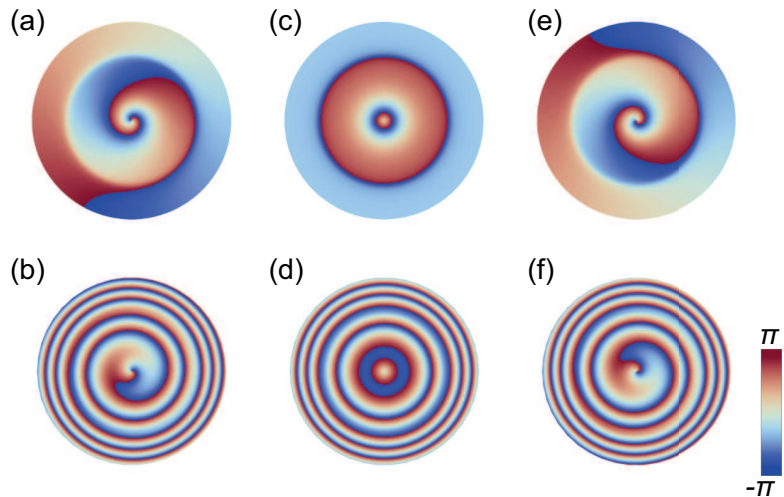}}
\caption{Examples of the phase components of the simulated polariton condensates with ($m=\pm1$) and without ($m=0$) azimuthal polariton currents.
Upper panels show phases of the concentric ring polariton condensates, lower panels show corresponding phases relative to the phase of the reference spherical wave.} \label{FIG_Simul}
\end{figure}

Figure~\ref{FIG_Density}(a) shows the simulated density distribution of the double ring polariton condensate. 
Weak submicrometer displacement of the pump spot hardly affects its shape, so it remains the same in all simulations discusses below.
We can judge the emergence of azimuthal polariton currents by the phase variation of the condensate around the micropillar.  
In the upper panels in Fig.~\ref{FIG_Simul} we show three examples of the phase distribution of the simulated concentric ring polariton condensates.
In panel (c) the phase does not vary in the azimuthal direction, while in (a) and (e) the phases change by $2\pi$ with one turn around the pillar in the counterclockwise and clockwise directions in both condensate rings, respectively. 
For quantifying polariton vortices, we use the winding number~$m$.
Following the definition, given in the Introduction, we obtain $m= +1, 0$ and~$-1$ for panels (a), (b) and (c), respectively.
The lower panels in Fig.~\ref{FIG_Simul} show the phases of the condensates relative to the phase of the spherical wave.
These figures are helpful for comparing the simulation results with the experimental observations discussed below.

\section{Experimental observation of double ring polariton vortices.}

Vortex and non-vortex states of double concentric ring polariton condenstes have been observed by us in the experiment.
We excited the polariton condensates in a cylindrical pillar of diameter of $30 \, \mu\text{m}$, etched in the GaAs microcavity with a set of embedded quantum wells.
The excitation was performed at temperature of 4 K by a nonresonant laser pump beam focused close to the center of the micropillar.
We measured photoluminescence from the condensate to reveal the density distribution of polaritons in the micropillar.
To get access to the phase of the polariton condensate, we measured interference of the condensate photoluminescence with the spherical reference wave using the Mach-Zehnder interferometer.
The spherical wave was obtained by magnifying a small peripheral area of the condensate image.
More details on the experimental setup and the sample are given in~\cite{PhysRevB97195149,LukoshkinPaper2022}.

\begin{figure}[htbp]
\centerline{\includegraphics[width=4in]{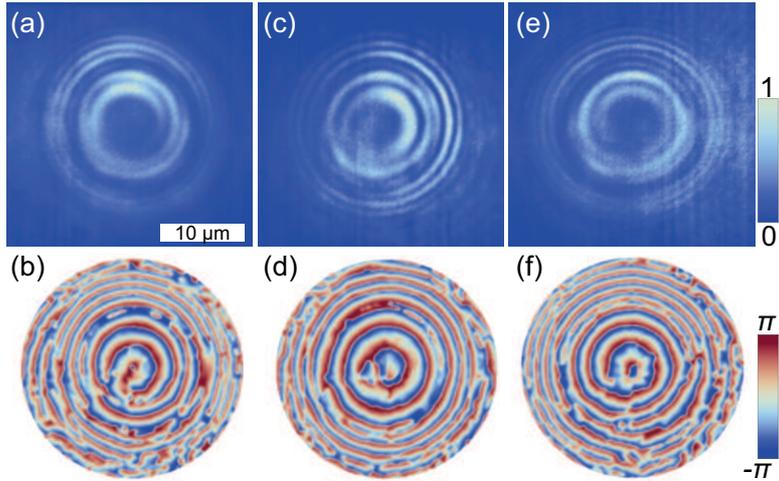}}
\caption{Observation of concentric ring polariton condensates with azimuthal polariton currents.
Upper panels show images of interference of the polariton condensates with the reference spherical wave.
Lower panels show corresponding distribution of the phases of the condensates relative to the phase of the reference wave.} \label{FIG_Exp}
\end{figure}

Figure~\ref{FIG_Density}(b) shows photoluminescence from the double ring exciton polariton condensate in the micropillar.
Upper panels in Fig.~\ref{FIG_Exp} show interferometry images of three observed double ring condensates.
The interference fringes in (a) and (e) have the shape of spirals, diverging in counterclockwise and clockwise directions, respectively.
The fringes in (c) represent concentric rings.
Brightness of the fringes fades beyond the condensate, including the slit between the condensate rings.

We use the Fourier-transform method adapted for closed interference fringes~\cite{PhysRevB97195149} for extracting phase of the condensate from the interferograms.
The lower panels in Fig.~\ref{FIG_Exp} show the phases of the observed polariton condensates relative to the phase of the spherical reference wave.
One can see that the phase patterns repeat the spiral or concentric ring shape of the interferometry images.
The comparison of the restored relative phases with the results of numerical simulations (lower panels in Fig.~\ref{FIG_Simul}) allows to conclude that among the observed condensates, the two of them, (a,b) and (e,f), are in the vortex states with $m=+1$ and $m=-1$, respectively, while the remaining state (c,d) is in the non-vortex state with~$m=0$.

The observed condensates with different vorticity were obtained by slightly displacing the pump spot near the center of the micropillar.
Along with the inevitable inhomogeneities of the localizing potential landscape, the displacement led to breaking the rotational symmetry of the system and selection of the preferred direction for azimuthal polariton currents.
We also would like to note that at the chosen pump power in the micropillar under consideration we hadn't observed the polariton condensates with the radial quantum number other than~$n=2$.
The vortex states were highly stable and existed as long as the optical excitation was present.
The double concentric ring shape of the condensates also did not change during their observation.

\section{Conclusion}
In the manuscript, we have analyzed the conditions for exciting concentric ring exciton polariton condensates in cylindrical micropillars under the nonresonant optical pumping.
We have shown that manipulating by the imaginary part of the trapping potential, responsible for the spatial distribution of gain and loss in the micropillar, one can select the radial quantization states of the polariton condensate.
Based on numerical analysis, we have predicted excitation of azimuthal polariton current states in concentric ring polariton condensates.
Our predictions have been supported by experimental observation of stable double ring polariton vortices.
The results of our study have prospects for application in developing new interferometric devices and in coding and transferring optical information using the orbital degree of freedom of light, as well as for implementation of double-qubit logic gates.

\section*{Acknowledgements}
This work was done under the support of the Saint-Petersburg State University (Grant No. 91182694).
A.K. and P.S. acknowledge the support of Westlake University, Project 041020100118 and Program 2018R01002 funded by Leading Innovative and Entrepreneur Team Introduction Program of Zhejiang Province of China.
A.K.~acknowledges support from the Moscow Institute of Physics and Technology under the Priority 2030 Strategic Academic Leadership Program.
Numerical simulations were supported by the RF Ministry of Science and Higher Education under Agreement No. 0635-2020-0013 and by the RF Presidential Grants for state support of young scientists No. MK-4729.2021.1.2 and No. MK-5318.2021.1.2.

\end{document}